%% file: main.tex
\documentclass[conference]{IEEEtran}
\IEEEoverridecommandlockouts

\usepackage{cite}

\usepackage{mathtools}
\usepackage{bm}

\usepackage[colorlinks=true,bookmarks=false,citecolor=black,urlcolor=black,linkcolor=black]{hyperref}

\usepackage{cite}

\usepackage{amsmath,amssymb,amsfonts}

\usepackage{graphicx}
\usepackage{textcomp}
\usepackage{xcolor}
\usepackage[textsize=tiny]{todonotes}
\usepackage{tikz}
\usepackage{booktabs}
\usepackage{stmaryrd}
\usepackage{braket}

\makeatletter
\newcommand\HUGE{\@setfontsize\Huge{40}{50}}
\makeatother

\def\BibTeX{{\rm B\kern-.05em{\sc i\kern-.025em b}\kern-.08em
    T\kern-.1667em\lower.7ex\hbox{E}\kern-.125emX}}

\input{KITColors}

\input{Utils/macro}

\pgfplotsset{compat=1.17}
\begin{document}

\title{Quantum CSS LDPC Codes based on Dyadic Matrices for Belief Propagation-based Decoding\thanks{The work of Alessio Baldelli was partially supported by Agenzia per la Cybersicurezza Nazionale (ACN) under the programme for promotion of XL cycle PhD research in cybersecurity (CUP I32B24001750005). \\
This work has received funding from the European Research Council (ERC) under the European Union’s Horizon 2020 research and innovation programme (grant agreement No. 101001899) and the German Federal Ministry of
Research, Technology and Space (BMFTR) within the project Open6GHub+ (grant
agreement 16KIS2405)}
}

\author{
\IEEEauthorblockN{
Alessio Baldelli$^\dagger$,
Massimo Battaglioni$^\dagger$,
Jonathan Mandelbaum$^\star$,
Sisi Miao$^\star$,
Laurent Schmalen$^\star$
}
\IEEEauthorblockA{
$^\dagger$\,\emph{Department of Information Engineering}, Università Politecnica delle Marche, Ancona 60131, Italy\\
Email: \texttt{a.baldelli@pm.univpm.it}, \texttt{m.battaglioni@univpm.it}
}
\IEEEauthorblockA{
$^\star$\,\emph{Communications Engineering Lab (CEL)}, Karlsruhe Institute of Technology (KIT), 76187 Karlsruhe, Germany\\
Email: \texttt{\{firstname.lastname\}@kit.edu}
}
}

\newacronym{AWGN}{AWGN}{additive white Gaussian noise}
\newacronym{BER}{BER}{bit error rate}
\newacronym{BP}{BP}{belief propagation}
\newacronym{BPSK}{BPSK}{binary phase-shift keying}
\newacronym{CAMEL}{CAMEL}{Cycle Assembling and Mitigating with EnsembLe decoding}
\newacronym{CSS}{CSS}{Calderbank--Shor--Steane}
\newacronym{DPM}{DPM}{dyadic permutation matrix}
\newacronym{LDPC}{LDPC}{low-density parity-check}
\newacronym{LLR}{LLR}{log-likelihood ratio}
\newacronym{QD}{QD}{quasi-dyadic}
\newacronym{QD-LDPC}{QD-LDPC}{quasi-dyadic low-density parity-check}
\newacronym{QLDPC}{QLDPC}{quantum low-density parity-check}
\newacronym{SNR}{SNR}{signal-to-noise ratio}
\newacronym{SC-LDPC}{SC-LDPC}{spatially coupled low-density parity-check}
\newacronym{TI}{TI}{time-invariant}
\newacronym
  [longplural={parity-check matrices},
   shortplural={PCMs}]
  {PCM}{PCM}{parity-check matrix}
\newacronym{QC}{QC}{quasi-cyclic}
\newacronym{QSC}{QSC}{quantum stabilizer code}
\newacronym{QEC}{QEC}{Quantum error correction}
\newacronym{LER}{LER}{logical error rate}
\newacronym{BP4}{BP4}{quaternary belief propagation}

\maketitle

\begin{abstract}
Quantum low-density parity-check (QLDPC) codes provide a practical balance between error-correction capability and implementation complexity in quantum error correction (QEC). 
In this paper, we propose an algebraic construction based on dyadic matrices for designing both classical and quantum LDPC codes. 
The method first generates classical binary quasi-dyadic LDPC codes whose Tanner graphs have girth~6. 
It is then extended to the Calderbank--Shor--Steane (CSS) framework, where the two component parity-check matrices are built to satisfy the compatibility condition required by the recently introduced CAMEL-ensemble quaternary belief propagation decoder. 
This compatibility condition ensures that all unavoidable cycles of length 4 are assembled in a single variable node, allowing the mitigation of their detrimental effects by decimating that variable node.
\\
\end{abstract}


\section{Introduction}
\label{introduction}
\input{Sections/introduction} 

\section{Notation and preliminaries}
\label{sec:preli}
\input{Sections/notation}

\section{Code design}
\label{sec:codecon}
\input{Sections/design}

\section{Numerical results}
\label{sec:montec}
\input{Sections/numerical_results}

\section{Conclusions}
\label{sec:conclusions}
\input{Sections/conclusions}

\end{document}

%% file: KITColors.tex
\definecolor{kit-green}{RGB}{0, 150, 130}
\colorlet{KITgreen}{kit-green}
\colorlet{kit-green100}{kit-green}
\colorlet{kit-green90}{kit-green!90!white}
\colorlet{kit-green80}{kit-green!80!white}
\colorlet{kit-green70}{kit-green!70!white}
\colorlet{kit-green60}{kit-green!60!white}
\colorlet{kit-green50}{kit-green!50!white}
\colorlet{kit-green40}{kit-green!40!white}
\colorlet{kit-green30}{kit-green!30!white}
\colorlet{kit-green25}{kit-green!25!white}
\colorlet{kit-green20}{kit-green!20!white}
\colorlet{kit-green15}{kit-green!15!white}
\colorlet{kit-green10}{kit-green!10!white}
\colorlet{kit-green5}{kit-green!5!white}

\definecolor{kit-blue}{RGB}{70, 100, 170}
\colorlet{KITblue}{kit-blue}
\colorlet{kit-blue100}{kit-blue}
\colorlet{kit-blue90}{kit-blue!90!white}
\colorlet{kit-blue80}{kit-blue!80!white}
\colorlet{kit-blue70}{kit-blue!70!white}
\colorlet{kit-blue60}{kit-blue!60!white}
\colorlet{kit-blue50}{kit-blue!50!white}
\colorlet{kit-blue40}{kit-blue!40!white}
\colorlet{kit-blue30}{kit-blue!30!white}
\colorlet{kit-blue25}{kit-blue!25!white}
\colorlet{kit-blue20}{kit-blue!20!white}
\colorlet{kit-blue15}{kit-blue!15!white}
\colorlet{kit-blue10}{kit-blue!10!white}
\colorlet{kit-blue5}{kit-blue!5!white}

\definecolor{kit-royalblue}{RGB}{0, 45, 76}
\colorlet{kit-royalblue100}{kit-royalblue}
\colorlet{kit-royalblue90}{kit-royalblue!90!white}
\colorlet{kit-royalblue80}{kit-royalblue!80!white}
\colorlet{kit-royalblue70}{kit-royalblue!70!white}
\colorlet{kit-royalblue60}{kit-royalblue!60!white}
\colorlet{kit-royalblue50}{kit-royalblue!50!white}
\colorlet{kit-royalblue40}{kit-royalblue!40!white}
\colorlet{kit-royalblue30}{kit-royalblue!30!white}
\colorlet{kit-royalblue25}{kit-royalblue!25!white}
\colorlet{kit-royalblue20}{kit-royalblue!20!white}
\colorlet{kit-royalblue15}{kit-royalblue!15!white}
\colorlet{kit-royalblue10}{kit-royalblue!10!white}
\colorlet{kit-royalblue5}{kit-royalblue!5!white}

\definecolor{kit-iceblue100}{RGB}{30, 53, 69}
\definecolor{kit-iceblue70}{RGB}{68, 94, 111}
\definecolor{kit-iceblue50}{RGB}{168, 185, 196}
\definecolor{kit-iceblue30}{RGB}{218, 225, 230}

\definecolor{kit-red}{RGB}{162, 34, 35}
\colorlet{KITred}{kit-red}
\colorlet{kit-red100}{kit-red}
\colorlet{kit-red90}{kit-red!90!white}
\colorlet{kit-red80}{kit-red!80!white}
\colorlet{kit-red70}{kit-red!70!white}
\colorlet{kit-red60}{kit-red!60!white}
\colorlet{kit-red50}{kit-red!50!white}
\colorlet{kit-red40}{kit-red!40!white}
\colorlet{kit-red30}{kit-red!30!white}
\colorlet{kit-red25}{kit-red!25!white}
\colorlet{kit-red20}{kit-red!20!white}
\colorlet{kit-red15}{kit-red!15!white}
\colorlet{kit-red10}{kit-red!10!white}
\colorlet{kit-red5}{kit-red!5!white}

\definecolor{kit-yellow}{RGB}{252, 229, 0}
\colorlet{KITyellow}{kit-yellow}
\colorlet{kit-yellow100}{kit-yellow}
\colorlet{kit-yellow90}{kit-yellow!90!white}
\colorlet{kit-yellow80}{kit-yellow!80!white}
\colorlet{kit-yellow70}{kit-yellow!70!white}
\colorlet{kit-yellow60}{kit-yellow!60!white}
\colorlet{kit-yellow50}{kit-yellow!50!white}
\colorlet{kit-yellow40}{kit-yellow!40!white}
\colorlet{kit-yellow30}{kit-yellow!30!white}
\colorlet{kit-yellow25}{kit-yellow!25!white}
\colorlet{kit-yellow20}{kit-yellow!20!white}
\colorlet{kit-yellow15}{kit-yellow!15!white}
\colorlet{kit-yellow10}{kit-yellow!10!white}
\colorlet{kit-yellow5}{kit-yellow!5!white}

\definecolor{kit-orange}{RGB}{223, 155, 27}
\colorlet{KITorange}{kit-orange}
\definecolor{kit-orange}{RGB}{223, 155, 27}
\colorlet{kit-orange100}{kit-orange}
\colorlet{kit-orange90}{kit-orange!90!white}
\colorlet{kit-orange80}{kit-orange!80!white}
\colorlet{kit-orange70}{kit-orange!70!white}
\colorlet{kit-orange60}{kit-orange!60!white}
\colorlet{kit-orange50}{kit-orange!50!white}
\colorlet{kit-orange40}{kit-orange!40!white}
\colorlet{kit-orange30}{kit-orange!30!white}
\colorlet{kit-orange25}{kit-orange!25!white}
\colorlet{kit-orange20}{kit-orange!20!white}
\colorlet{kit-orange15}{kit-orange!15!white}
\colorlet{kit-orange10}{kit-orange!10!white}
\colorlet{kit-orange5}{kit-orange!5!white}

\definecolor{kit-lightgreen}{RGB}{140, 182, 60}
\colorlet{KITlightgreen}{kit-lightgreen}
\colorlet{kit-lightgreen100}{kit-lightgreen}
\colorlet{kit-lightgreen90}{kit-lightgreen!90!white}
\colorlet{kit-lightgreen80}{kit-lightgreen!80!white}
\colorlet{kit-lightgreen70}{kit-lightgreen!70!white}
\colorlet{kit-lightgreen60}{kit-lightgreen!60!white}
\colorlet{kit-lightgreen50}{kit-lightgreen!50!white}
\colorlet{kit-lightgreen40}{kit-lightgreen!40!white}
\colorlet{kit-lightgreen30}{kit-lightgreen!30!white}
\colorlet{kit-lightgreen25}{kit-lightgreen!25!white}
\colorlet{kit-lightgreen20}{kit-lightgreen!20!white}
\colorlet{kit-lightgreen15}{kit-lightgreen!15!white}
\colorlet{kit-lightgreen10}{kit-lightgreen!10!white}
\colorlet{kit-lightgreen5}{kit-lightgreen!5!white}

\definecolor{kit-purple}{RGB}{163, 16, 124}
\colorlet{KITpurple}{kit-purple}
\colorlet{kit-purple100}{kit-purple}
\colorlet{kit-purple90}{kit-purple!90!white}
\colorlet{kit-purple80}{kit-purple!80!white}
\colorlet{kit-purple70}{kit-purple!70!white}
\colorlet{kit-purple60}{kit-purple!60!white}
\colorlet{kit-purple50}{kit-purple!50!white}
\colorlet{kit-purple40}{kit-purple!40!white}
\colorlet{kit-purple30}{kit-purple!30!white}
\colorlet{kit-purple25}{kit-purple!25!white}
\colorlet{kit-purple20}{kit-purple!20!white}
\colorlet{kit-purple15}{kit-purple!15!white}
\colorlet{kit-purple10}{kit-purple!10!white}
\colorlet{kit-purple5}{kit-purple!5!white}

\definecolor{kit-brown}{RGB}{167, 130, 46}
\colorlet{KITbrown}{kit-brown}
\colorlet{kit-brown100}{kit-brown}
\colorlet{kit-brown90}{kit-brown!90!white}
\colorlet{kit-brown80}{kit-brown!80!white}
\colorlet{kit-brown70}{kit-brown!70!white}
\colorlet{kit-brown60}{kit-brown!60!white}
\colorlet{kit-brown50}{kit-brown!50!white}
\colorlet{kit-brown40}{kit-brown!40!white}
\colorlet{kit-brown30}{kit-brown!30!white}
\colorlet{kit-brown25}{kit-brown!25!white}
\colorlet{kit-brown20}{kit-brown!20!white}
\colorlet{kit-brown15}{kit-brown!15!white}
\colorlet{kit-brown10}{kit-brown!10!white}
\colorlet{kit-brown5}{kit-brown!5!white}

\definecolor{kit-cyan}{RGB}{35, 161, 224}
\colorlet{KITcyan}{kit-cyan}
\colorlet{KITcyanblue}{kit-cyan}
\colorlet{kit-cyan100}{kit-cyan}
\colorlet{kit-cyan90}{kit-cyan!90!white}
\colorlet{kit-cyan80}{kit-cyan!80!white}
\colorlet{kit-cyan70}{kit-cyan!70!white}
\colorlet{kit-cyan60}{kit-cyan!60!white}
\colorlet{kit-cyan50}{kit-cyan!50!white}
\colorlet{kit-cyan40}{kit-cyan!40!white}
\colorlet{kit-cyan30}{kit-cyan!30!white}
\colorlet{kit-cyan25}{kit-cyan!25!white}
\colorlet{kit-cyan20}{kit-cyan!20!white}
\colorlet{kit-cyan15}{kit-cyan!15!white}
\colorlet{kit-cyan10}{kit-cyan!10!white}
\colorlet{kit-cyan5}{kit-cyan!5!white}

\definecolor{kit-gray}{RGB}{0, 0, 0}
\colorlet{KITgray}{kit-gray}
\colorlet{kit-gray100}{kit-gray}
\colorlet{kit-gray90}{kit-gray!90!white}
\colorlet{kit-gray80}{kit-gray!80!white}
\colorlet{kit-gray70}{kit-gray!70!white}
\colorlet{kit-gray60}{kit-gray!60!white}
\colorlet{kit-gray50}{kit-gray!50!white}
\colorlet{kit-gray40}{kit-gray!40!white}
\colorlet{kit-gray30}{kit-gray!30!white}
\colorlet{kit-gray25}{kit-gray!25!white}
\colorlet{kit-gray20}{kit-gray!20!white}
\colorlet{kit-gray15}{kit-gray!15!white}
\colorlet{kit-gray10}{kit-gray!10!white}
\colorlet{kit-gray5}{kit-gray!5!white}

\definecolor{KITpalegreen}{RGB}{130,190,60}
\colorlet{kit-maigreen100}{KITpalegreen}
\colorlet{kit-maigreen70}{KITpalegreen!70}
\colorlet{kit-maigreen50}{KITpalegreen!50}
\colorlet{kit-maigreen30}{KITpalegreen!30}
\colorlet{kit-maigreen15}{KITpalegreen!15}

%% file: Utils/macro.tex
\usepackage{comment}
\usepackage{theorem}
\usepackage{calligra}
\usepackage[acronym, shortcuts]{glossaries}
\usepackage[dvipsnames]{xcolor} %

\newcommand{\6}{\mathbf }

\usepackage{tikz}
\usetikzlibrary{arrows,automata, positioning}
\usetikzlibrary{calc, chains,
                fit,
                positioning,
                shapes}
\usetikzlibrary{patterns}       

\usepackage{pgfplots}
\pgfplotsset{compat=newest} %

\usetikzlibrary{matrix,fadings,decorations.pathmorphing,intersections}
\usepgfplotslibrary{fillbetween}
\pgfdeclarelayer{bg}
\pgfsetlayers{bg,main}
\makeatletter
\tikzset{use path/.code=\tikz@addmode{\pgfsyssoftpath@setcurrentpath#1}}

\definecolor{myblue}{RGB}{0,0,205}
\definecolor{mygreen}{RGB}{80,160,80}
\definecolor{myred}{RGB}{178,34,34}
\definecolor{mygray}{RGB}{211,211,211}
\definecolor{myyellow}{rgb}{1.0, 0.75, 0.0}
\definecolor{mygreen2}{rgb}{0.4, 0.69, 0.2}
\definecolor{myred2}{rgb}{0.89, 0.15, 0.21}
\definecolor{myblue2}{rgb}{0.0, 0.47, 0.75}

{\theorembodyfont{\rmfamily}
   }
{\theorembodyfont{\rmfamily}
   \newtheorem{The}{{\textbf Theorem}}}
{\theorembodyfont{\rmfamily}
   }
{\theorembodyfont{\rmfamily}
   \newtheorem{Lem}{{\textbf Lemma}}}
{\theorembodyfont{\rmfamily}
   }
{\theorembodyfont{\upshape}
   \newtheorem{Def}{{\textbf Definition}}}
{\theorembodyfont{\upshape}
   \newtheorem{Exa}{{\textbf Example}}}
{\theorembodyfont{\rmfamily}
   \newtheorem{Ass}{{\textbf Remark}}}

\usetikzlibrary{patterns}
\usepackage{scalerel}

%% file: Sections/introduction.tex
\ac{QEC} is a fundamental component in the realization of fault-tolerant quantum computation and communication systems. 
Due to the intrinsic fragility of quantum information, physical qubits are highly susceptible to noise and decoherence, making error control indispensable for reliable operation \cite{Shor1995, Steane1996, Calderbank1996, Gottesman1997}. 

The stabilizer formalism provides a powerful algebraic framework for \ac{QEC} by embedding $k$ logical qubits into a subspace of the $n$--qubit Hilbert space stabilized by a set of commuting Pauli operators \cite{Nielsen2010}. 
Within \acp{QSC}, the \ac{CSS} code family \cite{Calderbank1996, Steane1996,calderbank1998quantum} is widely used. In this family, the stabilizer structure decomposes into two classical binary codes $(C_X, C_Z)$ with \acp{PCM} $\6H_X$ and $\6H_Z$, respectively, satisfying the well-known \emph{orthogonality condition} $\6H_X\6H_Z^\top=0$. 

Among \ac{CSS}-type codes, \ac{QLDPC} codes, i.e., stabilizer codes represented by a sparse \ac{PCM}, have emerged as a promising class, 
which enables scalable and efficient \ac{BP} decoding, and a good trade-off between error correction capabilities and the depth of the associated quantum circuit \cite{Gallager, MacKay1999, Poulin2008}. However, classical \ac{LDPC} design principles do not directly translate into the quantum setting, where the orthogonality constraint between $\6H_X$ and $\6H_Z$ restricts the admissible code ensemble and complicates the elimination of short cycles in the Tanner graph \cite{Babar2015}.

Recent advances in structured \ac{QLDPC} code design have focused on preserving sparsity while ensuring algebraic regularity. 
Notable examples include quasi-cyclic constructions \cite{Hagiwara2007} and codes derived from projective geometries  \cite{Tang2024}. 
More recently, non-binary codes employing affine permutation matrices have been explored in \cite{Kasai2025}. In \cite{SisiCamel}, the \ac{CAMEL} framework was introduced. It employs \ac{CSS} codes in which all short cycles are concentrated around a single variable node, and it incorporates a decoding strategy that exploits this structure to neutralize their impact. The example codes used in \cite{SisiCamel} are based on circulant permutation matrices and finite geometries.

In this work, we extend the \ac{CAMEL} design philosophy by introducing a novel construction of \ac{QLDPC} codes based on dyadic matrices. Compared to the quasi-cyclic codes used in \cite{SisiCamel}, the proposed construction enables the design of new codes with higher rates and good decoding performance.
While dual-containing \ac{CSS} \ac{LDPC} codes based on dyadic matrices were recently introduced in \cite{baldelli_ISTC_2025}, their Tanner graphs exhibit a large number of length-$4$ cycles throughout the  \ac{PCM}, which limits their decoding performance. 
Adapting to the CAMEL framework, the newly proposed construction overcomes this limitation by exploiting affine-row dyadic structures to control the overlap between parity-check blocks, resulting in QLDPC codes suitable for quaternary \ac{BP} decoding.
Moreover, we show that the same principles can be leveraged to design classical \ac{QD-LDPC} codes \cite{Martinez2022, dyadics} with a girth of at least $6$.

The paper is organized as follows. Section~\ref{sec:preli} introduces the notation and background. Section~\ref{sec:codecon} presents the proposed dyadic-matrix–based classical and quantum \ac{LDPC} code constructions. Their error rate performance under ensemble \ac{BP} decoding over depolarizing channels is reported in Sec.~\ref {sec:montec}. Section~\ref{sec:conclusions} concludes the paper.

%% file: Sections/notation.tex
We denote the set of integers between $a$ and $b$, endpoints included, as $[a,b]$.
Let  $\mathbb F_{2^{\ell}}$ denote the finite field of order $2^{\ell}$, with $\ell \in \mathbb{N}$ and $\mathbb F_{2^\ell}^{\times}\coloneq \mathbb F_{2^\ell}\setminus\{0\}$. The element-wise XOR of binary vectors is denoted by $\oplus$, and the addition in $\mathbb F_{2^\ell}$ is denoted by $+$. Sums and multiplications of matrices are performed over $\mathbb{F}_2$.
The field $\mathbb F_{2^\ell}$ is constructed as $\mathbb F_2[x]/\langle \pi(x) \rangle$, 
where $\pi(x)$ is a primitive polynomial of degree~$\ell$, and its root $\alpha$ denotes a primitive element of the field. 
Each element  of the field can be written as 
\[
 \sum_{k=0}^{\ell-1} c_k \alpha^k, \qquad c_k \in \mathbb F_2.
\]

Matrices are denoted by bold uppercase letters, e.g., $\6H$, $\6P$;  row vectors are denoted by bold lowercase letters, e.g.,~$\6a$. 
The \emph{Hamming weight} (or \emph{weight}) of a binary vector $\6a$ is indicated with $|\6a|$, and it corresponds to the number of its non-zero entries. The \emph{support} of the vector $\6a$ consists of the set of indices of its non-zero entries.
Moreover, scalars in $\mathbb F_{2^{\ell}}$ are denoted by lowercase letters. The $m\times n$ all-zero and all-one matrices are $\60_{m\times n}$ and $\61_{m\times n}$, respectively. The transpose operator is $(\cdot)^\top$.

Given an integer $\ell\geq 1$, we define the \emph{ring of dyadic matrices} $\mathcal M_{\ell}(\mathbb F_2)$ as the set of $2^\ell\times 2^\ell$ matrices with entries over $\mathbb F_2$, structured as follows
    \begin{equation}
    \label{eq:dyadics}
        \6M = \begin{bmatrix}
        \6A & \6B \\
        \6B & \6A
        \end{bmatrix}, \quad \6A,\6B\in\mathcal M_{\ell-1}(\mathbb F_2).
    \end{equation}
For ${\ell = 0}$, ${\mathcal M_0(\mathbb F_2)\coloneq \mathbb F_2}$.
For any $\ell$, when equipped with standard matrix sum and multiplication, $\mathcal M_\ell(\mathbb F_2)$ forms a commutative ring. 
It is possible to show that any dyadic matrix is fully represented by its first row, called \emph{signature} \cite{santini2022reproducible}. 
Let ${\mathcal D_\ell(\mathbb F_2) = \{ \6D^{(0)}\!\coloneq \6I_{2^\ell\times2^\ell}, \6D^{(1)},\ldots, \6D^{(2^\ell-1)} \}\subseteq \mathcal M_\ell(\mathbb F_2)}$ be the set containing all dyadic matrices whose signatures have weight $1$. 
Then, each $\6D^{(i)}$ is called \emph{\ac{DPM}}.
Furthermore, a matrix that is built using only dyadic blocks is called \emph{\ac{QD}}. A code defined by a \ac{QD} generator matrix or by a \ac{QD} \ac{PCM} is naturally called \ac{QD}.

\subsection{Classical and Quantum Codes}
An $[n, k, d]$ classical binary linear code $C \subseteq \mathbb{F}_2^{n}$ has code length $n$, dimension $k$, 
and minimum Hamming distance $d$. 
The elements $\mathbf{c}$ of $C$ are the codewords, and ${d\coloneq \min_{\6c\in C\setminus\{\60\}} |\6c|}$.
The code rate is defined as $R = k/n$.
A code can be represented as the null space of a  \ac{PCM} $\6H\in\mathbb{F}_2^{r \times n}$, such that $\6H\6c^{\top} = \mathbf{0}, \, \, \forall \, \6c \in C$. The code redundancy is given by $\text{rank}(\6H)$. 
If $\6H$ is \emph{sparse}, i.e., the majority of its entries are zero, then the associated code is deemed to be an \ac{LDPC} code. Every \ac{PCM} can be interpreted as the adjacency matrix of a bipartite Tanner graph \cite{Tanner1981}, characterized by $n$ variable nodes and $r$ check nodes. The edges that link each other such nodes are represented by the non-zero entries in $\6H$. A closed path in the Tanner graph is called \emph{cycle}; we define the \emph{girth} $g$ to be the length of the shortest cycle in such a graph.

\acp{QSC}~\cite{Gottesman1997} are the quantum equivalent of classical binary linear codes. An $\llbracket n, k, d \rrbracket$ \ac{QSC}  $\mathcal{C} \subseteq (\mathbb{C}^2)^{\otimes n}$ is called a stabilizer code because it is stabilized by all stabilizer generators $s \in \mathcal{S}$ of the code\footnote{To distinguish between classical codes and \ac{QSC}, we denote the latter by the symbol $\mathcal{C}$ and the double-bracket notation $\llbracket \cdot \rrbracket$.}.
An $\llbracket n, k_1 - k_2, d \rrbracket$ \ac{CSS} code $\mathcal{C}$ is a special case of a \ac{QSC} \cite{Calderbank1996, Steane1996,calderbank1998quantum},  which can be constructed as follows. 
Let us consider two binary classical codes $C_1$ and $C_2$
with parameters $[n, k_1, d_1]$ and $[n, k_2, d_2]$, respectively, and such that $C_2 \subseteq C_1$. Given $\6{H}_X \in \mathbb{F}_2^{(n - k_1)\times n}$, and $\6{H}_Z \in \mathbb{F}_2^{k_2 \times n}$, which are the \acp{PCM} of $C_1$ and $C_2^{\perp}$, respectively, that satisfy the \emph{symplectic product} (also called \emph{orthogonality}) \emph{condition}
\begin{equation*}
    \6H_X \6H_Z^\top = \60 \quad (\text{or } \6H_Z \6H_X^\top = \60),
\end{equation*}
then, by using $\mathbb{F}_4=\{0,1,\omega:=\alpha,\bar{\omega}=\alpha^2\}$, the associated \ac{CSS} code $\mathcal{C}$ is represented by the quaternary \ac{PCM}
\begin{equation} \label{eq:CSS_codes}
        \begin{bmatrix}
            \omega \6H_{X}  \\ 
            \bar{\omega}\6{H}_Z 
        \end{bmatrix}.
\end{equation}
The minimum distance $d$ is lower-bounded by $\text{min} \{ d_1, d_2 \}$, and the rate of $\mathcal{C}$ is $R_Q = (k_1-k_2)/n$.

\subsection{CAMEL}

Up to this day, several constructions for the binary \acp{PCM} $\6H_X$ and $\6H_Z$ of a \ac{CSS} code guarantee a girth of at least $6$ for each of the respective associated Tanner graphs \cite{Hagiwara2007, Kasai2025, pacenti_margulis}.
However, when considering quaternary \ac{BP} decoding, which is favorable compared to binary \ac{BP} decoding because it takes correlation between $X$ and $Z$ errors into account, the performance may be degraded by inevitable $4$‑cycles in the Tanner graph of the quaternary \ac{PCM} $\6H$ \cite{SisiCamel}. 
The \ac{CAMEL} framework introduced in \cite{SisiCamel} jointly designs the code and quaternary \ac{BP}-based decoder to alleviate the harmful effect of these short cycles.

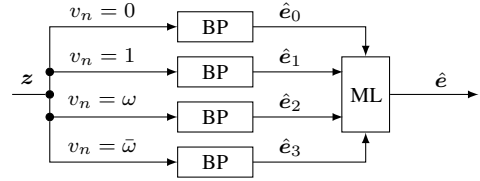
\begin{figure}
    \centering
    \input{Figures/block_diagramm}
    \caption{Block diagram of the CAMEL decoder according to \cite{SisiCamel}.}
    \label{fig:block_diagramm}
\end{figure}

\begin{Def}
A CAMEL decoder and code pair consists of the ensemble decoder depicted in Fig.~\ref{fig:block_diagramm} together with a CSS code represented by binary \acp{PCM}
\begin{equation}
  \6H_X=
\begin{pmatrix}\6H'_X&\61_{r\times1}
\end{pmatrix}
,\quad 
\6H_Z=
\begin{pmatrix}\6H'_Z&\61_{r \times 1}
\end{pmatrix},  
\label{eq:allone}
\end{equation}
where $\6H'_X,\6H'_Z\in \mathbb{F}_2^{r\times n}$ and the submatrices satisfy the \emph{CAMEL condition} 
\begin{equation}\label{eq:camel_condition}
    \6H'_X (\6H'_Z)^\top
    = \61_{r \times r}.
\end{equation}
\end{Def}

In \cite{SisiCamel}, two construction methods are presented that ensure a girth of at least $6$ for the Tanner graphs associated with $\6H'_X$ and $\6H'_Z$, respectively.
In this case, all (unavoidable) cycles of length $4$ in the Tanner graph associated with $\6H$ are assembled in the last column.
By guessing all distinct values of the last variable node, as done in each path of the ensemble decoder in Fig.~\ref{fig:block_diagramm},  the influence of the cycles of length $4$ is effectively suppressed. The final estimate is obtained by choosing the error estimate with the lowest weight satisfying the syndrome among all decoding paths.

\subsection{\acp{PCM}  based on \acp{DPM}}
We represent each \ac{QD} matrix based on \acp{DPM} by a special \emph{exponent matrix} 
\(\6P \in \mathbb{F}_{2^\ell}^{w \times N}\), where each entry 
\(p_{i,j} \in \mathbb{F}_{2^\ell}\) selects one \ac{DPM} 
from the set \(\mathcal{D}_\ell(\mathbb{F}_2)\) through a fixed bijection 
\(\psi : \mathbb{F}_{2^\ell} \to [0,2^\ell\!-\!1]\), defined as
\[
\psi\!\left(\sum_{k=0}^{\ell-1} c_k \alpha^k\right)
= \sum_{k=0}^{\ell-1} c_k 2^k,
\] 
and we will denote by $\lambda_j \coloneqq \psi^{-1}(j) \in \mathbb F_{{2}^\ell}$ the field element corresponding to integer $j$. Specifically, the \ac{DPM} in position \((i,j)\) is given by $\6D^{(\psi(p_{i,j}))}$,
where each \(\6D^{(l)} \in \mathcal{D}_\ell(\mathbb{F}_2)\) has a single~$1$ 
in the $l$-th position of its first row, with counting starting from $0$. 
The operation of substituting the entries of $\6P$ with such \acp{DPM} to obtain the \ac{PCM} $\6H$ is called \emph{lifting}.  
The binary representation $\rho_{i,j}\in\mathbb{F}_2^{\ell}$ of $p_{i,j} \in \mathbb{F}_{{2}^\ell}$ is obtained via the fixed vector space isomorphism $\varphi:\mathbb F_{2^\ell}\to\mathbb{F}_2^{\ell}$ with
\[
\varphi\!\left(\sum_{k=0}^{\ell-1} c_k \alpha^k\right)
\coloneq (c_{\ell-1}, \ldots, c_0),
\]
so that 
\[
\varphi(p_{i,j}+p_{i',j'})
= \varphi(p_{i,j})\oplus\varphi(p_{i',j'})
= \rho_{i,j}\oplus\rho_{i',j'}.
\]

We refer to a \emph{block-row} (\emph{column-row})  as any row (column) of $\6H$ composed exclusively of dyadic blocks.

%% file: Figures/block_diagramm.tex
		\tikzset{line/.style={-latex}} 
		\begin{tikzpicture}
    \node[draw=none] at (0,3.2) {\footnotesize$\bm{z} $};
    \node[draw,circle,fill,inner sep=1pt] at (0.3,3)(s) {};
    \node[draw,circle,fill,inner sep=1pt] at (0.3,3.3) (s1){};
    \node[draw,circle,fill,inner sep=1pt] at (0.3,2.7) (s2) {};

    \draw[] (-0.2,3) -- (s);
    
    \draw[] (2,3.7) rectangle ++(1,0.4);
    \node[draw=none] at (1,4.1) {\footnotesize $v_n=0$};
    \node[draw=none] at (2.5,3.9)(BP1) {\footnotesize BP};  
    \draw[line] (s1) |- (2,3.9);
    
    \draw[] (2,3.1) rectangle ++(1,0.4);
    \node[draw=none] at (1,3.5) {\footnotesize $v_n=1$};
    \node[draw=none] at (2.5,3.3)(BP2) {\footnotesize BP};
    \draw[line] (s) |- (2,3.3);
    
    \draw[] (2,2.5) rectangle ++(1,0.4);
    \node[draw=none] at (1,2.9) {\footnotesize $v_n=\omega$};
    \node[draw=none] at (2.5,2.7)(BP3) {\footnotesize BP};
    \draw[line] (s) |- (2,2.7);
    \draw[] (2,1.9) rectangle ++(1,0.4);
    \node[draw=none] at (1,2.3) {\footnotesize $v_n=\bar{\omega}$};
    \node[draw=none] at (2.5,2.1)(BP4) {\footnotesize BP};
    \draw[line] (s2) |- (2,2.1);

    \node[draw,rectangle,text width=0.4cm, minimum height=1cm,align=center] at (4.5,3)(w) {\footnotesize ML};

    \node[draw=none] at (3.5,4.1) {\footnotesize $\hat{\bm{e}}_0$};
    \node[draw=none] at (3.5,3.5) {\footnotesize $\hat{\bm{e}}_1$};
    \node[draw=none] at (3.5,2.9) {\footnotesize $\hat{\bm{e}}_2$};
    \node[draw=none] at (3.5,2.3) {\footnotesize $\hat{\bm{e}}_3$};

    \draw[line] (3,3.9) -| (w);
    \draw[line] (3,3.3) -- (4.2,3.3);
    \draw[line] (3,2.7) -- (4.2,2.7);
    \draw[line] (3,2.1) -| (w);
    
     \node[draw=none] at (5.5,3.2) {\footnotesize$\hat{\bm{e}}$};
     \draw[line] (w) -- (6,3);
     \end{tikzpicture}

%% file: Sections/design.tex
In the following, we introduce our code construction both for classical and quantum codes.

\subsection{Classical QD-LDPC Codes Design with Girth~$6$}\label{subsec:class}

Let us consider classical \ac{QD-LDPC} codes with ${N=L \coloneqq 2^{\ell}}$, ${3 \le w \le 2^\ell}$. Our goal is to design an exponent matrix $\6P$ corresponding to a \ac{QD-LDPC} code with girth $6$. To this end, we propose to obtain each row of $\6P$ as an affine permutation of the field elements of $\mathbb F_{2^\ell}$, i.e.,
\begin{equation}
    p_{u,j} \coloneqq a_u \lambda_j + b_u, 
    \label{eq:affine_generation}
\end{equation}
where $u\in[0, \, w-1], j \in [0,2^\ell-1]$, and $\lambda_j = \psi^{-1}(j) \in \mathbb F_{{2}^\ell}$. Then, the set of ${a_u \in \mathbb F_{2^\ell}^{\times}}$ are called \emph{multipliers} which are all distinct. Additionally, $b_u \in \mathbb F_{2^\ell}$.
Each distinct multiplier $a_u$ defines a unique affine permutation of $\mathbb F_{2^\ell}$. We will show that this rule yields \ac{LDPC} matrices with Tanner graphs of girth~$6$, if the multipliers within the same exponent matrix are distinct. %

\begin{Lem} \cite[Theorem 3]{baldelli_ISTC_2025}
Let $\6P \in \mathbb F_{2^{\ell}}^{w\times N}$ be the exponent matrix of a \ac{QD-LDPC} code. By applying the lifting procedure on $\6P$ using the associated  \acp{DPM} of size $2^\ell\times 2^\ell$, we obtain the Tanner graph of the corresponding binary \ac{PCM} $\6H$ which contains at least $\ell$ cycles of length $4$ if 
\begin{equation}
\label{eq:4cycle-condition}
\rho_{i_1,j_1}\oplus \rho_{i_2,j_1} \;=\; \rho_{i_2,j_2}\oplus \rho_{i_1,j_2},
\end{equation}
with $i_1\neq i_2$ and $j_1\neq j_2$.
\label{lem:4cyc}
\end{Lem}

\begin{The}
\label{the:girth6}
Let $\6P \in \mathbb F_{2^\ell}^{w\times N}$ be an exponent matrix generated according to \eqref{eq:affine_generation} with distinct (non-zero) multipliers $a_1,\dots, a_w$.
Then, for every pair of distinct rows ${u,v \in [0,w-1]}$, the mapping
\begin{equation}
    j \mapsto p_{u,j} + p_{v,j} = (a_u + a_v)\lambda_j + (b_u + b_v), \;\;\; j\in[0,2^\ell-1],
\end{equation}
with $\lambda_j=\psi^{-1}(j)$, yields a distinct permutation of $\mathbb F_{2^\ell}$.
Consequently, the Tanner graph of $\6H$ obtained by lifting $\6P$ has girth~$6$.
\end{The}
\begin{IEEEproof}
Since $a_u \neq a_v$, we have $a_u + a_v \in \mathbb F_{2^\ell}^{\times}$. Hence, the coefficient $(a_u + a_v)$ has a multiplicative inverse.
Thus, using the same argument as below, it is easy to show that the map ${j \mapsto (a_u + a_v)\lambda_j + (b_u + b_v)}$ is bijective over $\mathbb F_{2^\ell}$, implying that all values $p_{u,j} + p_{v,j}$ are distinct for ${j\in [0,2^\ell-1]}$. 

Let again $\rho_{u,j}$ denote the $\ell$-bit binary representation of $p_{u,j}$ through 
$\varphi$.
Assume two columns with indices $j_1\neq j_2$ satisfy
\[
\rho_{u,j_1}\oplus\rho_{v,j_1} = \rho_{u,j_2}\oplus\rho_{v,j_2}.
\]
Then, applying $\varphi^{-1}$ yields
\[
p_{u,j_1} + p_{v,j_1} = p_{u,j_2} + p_{v,j_2}.
\]
Substituting the affine form \eqref{eq:affine_generation} gives
\[
(a_u+a_v)(\lambda_{j_1}-\lambda_{j_2}) = 0.
\]
Since ${(a_u+a_v)\neq 0}$, we must have $\lambda_{j_1}=\lambda_{j_2}$ resulting in a contradiction.
Hence, all  $\rho_{u,j_1}\oplus\rho_{v,j_1}$ are distinct for any pair of rows, and no 4-cycles exist according to Lemma \ref{lem:4cyc}.
\end{IEEEproof}

\subsection{Quantum LDPC Code Design}

In this section, we propose the construction of \ac{QD} \ac{CSS} codes satisfying the CAMEL condition \eqref{eq:camel_condition}. 
To this end, we define two \ac{QD} matrices $\6H'_X$ and $\6H'_Z$ with corresponding exponent matrices $\6P'_X$ and $\6P'_Z$.

\begin{Ass}
Let $\6{r}_X$ be a binary row of $\6H'_X$ and $\6{r}_Z$ be a binary row of $\6H'_Z$. The inner product
$\6{r}_X \cdot \6{r}_Z^\top$ is taken over $\mathbb{F}_{2}$.
Hence, it equals the \emph{parity} of the number of positions where both rows have a $1$. Therefore, stating that $\6H'_X \left(\6H'_Z\right)^\top$ has all entries equal to $1$ over $\mathbb F_{2}$ means that for \emph{every} pair of rows---one from $\6H'_X$ and one from $\6H'_Z$---the number of positions where both entries are $1$ is odd.
\end{Ass}

As in Sec. \ref{subsec:class}, each row of  $\6P'_X$ and $\6P'_Z$ is generated through an affine
transformation in $\mathbb F_{2^\ell}$, that is,
\begin{equation}
    p^{(M)}_{u,j} = a^{(M)}_u \lambda_j + b^{(M)}_u
    \label{eq:affine-again}
\end{equation}
where $a^{(M)}_u \in \mathbb F_{2^\ell}^{\times}$, $
           b^{(M)}_u \in\mathbb F_{2^\ell}$, $j\in [0,2^\ell-1]$, $\lambda_j=\psi^{-1}(j)$, and $M \in \{X, Z\}$. Also, $N=L=2^{\ell}$ ensures a bijection between the blocks forming the \acp{PCM} and the $\mathbb F_{2^\ell}$ elements.

\begin{Lem}
   Let two exponent matrices, which we denote as $\6P'_X,\6P'_Z\in \mathbb F_{2^\ell}^{w\times N}$, have affine rows
as per \eqref{eq:affine-again}, 
with ${a^{(X)}_u\neq a^{(Z)}_v}$, $\forall \, u,v$.
We form  $\6H'_X,\6H'_Z$ by lifting $\6P'_X$ and $\6P'_Z$ with the corresponding \acp{DPM}. 
Then, \emph{every} pair consisting of one row from the $u$th block-row of $\6H'_X$ and one row from the $v$th block-row of $\6H'_Z$ has exactly one overlapping non-zero position.
\label{lem:1overlap}
\end{Lem}
\vspace*{-0.5em}
\begin{IEEEproof}
It is convenient to label the $2^\ell$ lifted rows within the $u$th block-row of $\6H'_X$ by $r\in [0,2^\ell-1]$ and within the $v$th block-row of $\6H'_Z$ by $s\in [0,2^\ell-1]$. We also label the columns within each block-column with elements of $[0,2^\ell-1]$. 
By the definition of the \ac{DPM}, in block-column $j$, row $r$ has its unique $1$ at column $$\psi\left(\varphi^{-1}(\varphi(p_{u,j}^{(X)})\oplus\varphi(\lambda_r))\right)=\psi\left(p_{u,j}^{(X)}+\lambda_r\right),$$ whereas row $s$ has its unique $1$ at column
$$\psi\left(\varphi^{-1}(\varphi(p_{v,j}^{(Z)})\oplus\varphi(\lambda_s))\right)=\psi\left(p_{v,j}^{(Z)}+\lambda_s\right).$$
Therefore, in block-column $j$, the two lifted rows overlap iff\footnote{All operations are over $\mathbb F_{2^\ell}$. 
Since the field has characteristic~2. 
The minus sign is retained only for the sake of readability.}
\[
\lambda_r+p^{(X)}_{u,j}=\lambda_s+p^{(Z)}_{v,j}
\;\Longleftrightarrow\;
p^{(X)}_{u,j}-p^{(Z)}_{v,j}=\lambda_s-\lambda_r.
\]
Define the difference map
\[
\Delta(j)\coloneq p^{(X)}_{u,j}-p^{(Z)}_{v,j}
=\bigl(a^{(X)}_u-a^{(Z)}_v\bigr)\lambda_j+\bigl(b^{(X)}_u-b^{(Z)}_v\bigr).
\]
Since $a^{(X)}_u\neq a^{(Z)}_v$, the coefficient of $\lambda_j$ is nonzero, and so $\Delta:\mathbb F_{2^\ell}\!\to\!\mathbb F_{2^\ell}$ is an affine bijection. 
Given the pair of lifted-row indices $(r,s)$, set $\delta\coloneq\psi( \lambda_s-\lambda_r)$. 
By bijectivity, there exists a unique block index $j^\star$ such that $\Delta(j^\star)=\lambda_\delta$, i.e.,
\[
p^{(X)}_{u,j^\star}-p^{(Z)}_{v,j^\star}=\lambda_s-\lambda_r.
\]
Hence the two rows overlap \emph{only} in block $j^\star$. 
Inside that block, each of the two rows has a single $1$, and the equality above forces these to occur in the same column, yielding exactly one common $1$ overall. 
No other block can satisfy the equality because $j^\star$ is unique. 
Thus, any such pair of lifted rows has exactly one overlap.
\end{IEEEproof}

\begin{The}
\label{thm:camel-condition}
Let $\6P'_X, \6P'_Z \in \mathbb F_{2^\ell}^{w\times N}$ be defined as in \eqref{eq:affine-again}. Assume all non-zero multipliers $a^{(M)}_u$ are distinct both within and across the two matrices, i.e.,
\[
\begin{aligned}
    & a^{(X)}_u \neq a^{(X)}_{u'} , && \text{for all } u \neq u', \\[2mm]
    & a^{(Z)}_v \neq a^{(Z)}_{v'} , && \text{for all } v \neq v', \\[2mm]
    & a^{(X)}_u \neq a^{(Z)}_v ,   && \text{for all } u,v.
\end{aligned}
\]
Then, the CAMEL condition \eqref{eq:camel_condition} is satisfied regardless of the choice of $b_u$.
\end{The}

\begin{IEEEproof}
Fix any $u$ in $\6P'_X$ and any $v$ in $\6P'_Z$. By the across-matrix distinctness, we have that $a^{(X)}_u\neq a^{(Z)}_v$.
Let us form $\6H'_X,\6H'_Z$ by lifting $\6P'_X$ and $\6P'_Z$ with the corresponding \acp{DPM}. 
By considering Lemma \ref{lem:1overlap}, for every choice of row indices $r,s\in\mathbb [0,2^\ell-1]$, the $r$th binary row in the $u$th block-row of $\6H'_X$ and the $s$th binary row in the $v$th block-row of $\6H'_Z$ overlap in \emph{exactly one} position (and in no other). 
Equivalently, their Hamming inner product over $\mathbb F_2$ equals $1$. Since $u$ and $v$, but also $r$ and $s$, are arbitrary, every cross inner product between a row of $\6H'_X$ and a row of $\6H'_Z$ equals $1$, which means all entries of $\6H'_X\left(\6H'_Z\right)^\top$ are $1$ in $\mathbb F_2$. 
Therefore $\6H'_X\left(\6H'_Z\right)^\top=\61_{(n-k_1) \times k_2}$, i.e., satisfying \eqref{eq:camel_condition}.
\end{IEEEproof}

Finally, $\6H_X$ and $\6H_Z$ are obtained from $\6H'_X$ and $\6H'_Z$ as in \eqref{eq:allone}. Next, we provide an example.

\vspace*{-0.3em}

\begin{Exa}
   Let $w = 3$ and $\ell = 3$. Following~\eqref{eq:affine-again}, we choose the affine coefficients over the finite field
\(
\mathbb{F}_{2^{\ell}} = \mathbb{F}_{8} = \mathbb{F}_{2}[x]\big/\!\langle x^{3} + x + 1\rangle.
\) The coefficients $a^{(M)}_u$ are selected from the multiplicative group 
$\mathbb{F}_{8}^{\!\times}$, while $b^{(M)}_u \in \mathbb{F}_{8}$. Namely, 
    \[
\begin{array}{ll}
(a^{(X)}_0,b^{(X)}_0)=(\alpha,1), & (a^{(Z)}_0,b^{(Z)}_0)=(1,\alpha),\\[3pt]
(a^{(X)}_1,b^{(X)}_1)=(\alpha^2,\alpha^2), & (a^{(Z)}_1,b^{(Z)}_1)=(\alpha^3,\alpha^6),\\[3pt]
(a^{(X)}_2,b^{(X)}_2)=(\alpha^4,\alpha^4), & (a^{(Z)}_2,b^{(Z)}_2)=(\alpha^6,1).
\end{array}
\]
The corresponding exponent matrices are
\[
\6P'_X=
\begin{bmatrix}
1 & \alpha^{3} & \alpha^{6} & \alpha^{5} & \alpha & 0          & \alpha^{4} & \alpha^{2} \\
\alpha^{2} & 0          & \alpha^{5} & \alpha^{3} & \alpha & \alpha^{4} & 1 & \alpha^{6} \\
\alpha^{4} & 0          & 1 & \alpha^{5} & \alpha^{3} & \alpha^{6} & \alpha^{2} & \alpha
\end{bmatrix},
\]
\[
\6P'_Z=
\begin{bmatrix}
\alpha & \alpha^{3} & 0          & 1 & \alpha^{4} & \alpha^{5} & \alpha^{2} & \alpha^{6} \\
\alpha^{6} & \alpha^{4} & \alpha^{3} & 0          & \alpha & 1 & \alpha^{2} & \alpha^{5} \\
1 & \alpha^{2} & 0          & \alpha^{6} & \alpha^{3} & \alpha^{4} & \alpha & \alpha^{5}
\end{bmatrix}.
\]
By applying the standard lifting procedure with \acp{DPM} to the exponent matrices above, we obtain $\6H'_X$ and $\6H'_Z$, and append horizontally to both of them an all-one column yielding the \acp{PCM} $\6H_X$ and $\6H_Z$ for the \ac{CSS} framework.

\end{Exa}

%% file: Sections/numerical_results.tex
We assess the performance of quantum \ac{QD-LDPC} codes in terms of \ac{LER} through Monte Carlo simulations over the depolarizing channel, with a depolarizing probability $p$. 
We employ the CAMEL decoder described in Fig.~\ref{fig:block_diagramm}, 
which runs for at most $15$ iterations, and the simulation continues with the same $p$ until $100$ logical errors are detected.

Based on Theorem~\ref{thm:camel-condition}, we construct \ac{QLDPC} codes considering different underlying fields, i.e., we use $\ell\in \{4,5\}$, resulting in fields  $\mathbb{F}_{16}$ and $\mathbb{F}_{32}$, respectively. To construct QD-QLDPC codes, we simply use the largest allowed set of multipliers $a$ that are distinct and non-zero, and allow $\6H_X$ and $\6H_Z$ to have the same size. Therefore, one element of \( \mathbb{F}_{2^{\ell}}^{\times} \) is randomly discarded, leaving an even number \(2^{\ell}-2\) of field elements, which form the set of multipliers.
We have observed through extensive numerical simulations that, as expected, the choice of the discarded element does not affect the decoding performance. Therefore, with $\alpha$ being a primitive element of the underlying field, we can use,  without loss of generality, ${\mathcal{A}_X\coloneq\{1,\alpha, \alpha^2, \ldots, \alpha^{2^{\ell-1}-2}\}}$ and ${\mathcal{A}_Z\coloneq\{\alpha^{2^{\ell-1}-1}, \alpha^{2^{\ell-1}}, \ldots, \alpha^{2^{\ell}-2}\}}$ as the set of multipliers $a$ for $\6H_X$ and $\6H_Z$, respectively. Here, ${|\mathcal{A}_X|=|\mathcal{A}_Z| = 2^{\ell-1}-1}$. We always set $b=0$, and apply  the dyadic lifting, obtaining $\6H'_X, \6H'_Z \in \mathbb{F}_2^{2^{\ell}(2^{\ell-1}-1)\times 2^{2\ell}}$. Appending an all-one column as in \eqref{eq:allone} results in $\6H_X, \6H_Z \in \mathbb{F}_2^{2^{\ell}(2^{\ell-1}-1)\times (2^{2\ell}+1)}$. The parameters of the resulting  codes D1 and D2 are summarized in Table~\ref{tab:FG_codes}.

\begin{table}[t]
    \centering
    \caption{QD QLDPC codes constructed using Theorem~\ref{thm:camel-condition}.}
    \label{tab:FG_codes}
    \begin{tabular}{cccccc}
    \toprule
         Code&$n$&$k$&$R_Q$&$\ell$ \\
         \hline
         D1&257&121&0.47&4   \\
         D2&1025&583&0.57&5  \\ 
         \bottomrule
    \end{tabular}
\end{table}

\begin{table}[t]
    \centering
    \caption{Parameters of the \ac{QC} codes C1 and C2 constructed according to~\cite{SisiCamel}, the Euclidean geometry code E5 from~\cite{SisiCamel}, and the bicycle code B1.}
    \label{tab:reference_codes}
    \begin{tabular}{ccccc}
    \toprule
         Code&$n$&$k$&$R_Q$ \\
         \hline
         C1&290&128&0.44 \\
         C2&962&540&0.56 \\
         E5&1057&571&0.54 \\
         B1&800&400&0.5 \\
         \bottomrule
    \end{tabular}
    \vspace{-1em}
\end{table}

\begin{figure}
    \centering
        \vspace{-1em}
    \input{Figures/FER_D1}
        \caption{\ac{LER} of \ac{QLDPC} codes as a function of the depolarizing probability~$p$.}
    \label{fig:fer_d1}
\end{figure}
\begin{figure}[t]
    \centering
    \input{Figures/FER_codes_from_theorem2}
    \caption{\ac{LER} of \ac{QLDPC} codes as a function of the depolarizing probability~$p$.}
     \vspace{-1em}
    \label{fig:first_CAMEL}
\end{figure}

In Figs. \ref{fig:fer_d1} and \ref{fig:first_CAMEL}, we report the post-decoding \ac{LER} of the constructed codes over a depolarizing channel where the three Pauli errors occur with equal probability $p/3$. All BP decoders employ  sum-product \ac{BP} decoding with a flooding-schedule. We consider three decoding strategies: plain quaternary \ac{BP} decoding; a genie-aided quaternary \ac{BP} decoder, in which the correct error value of the last qubit is provided to the decoder; and the proposed 4-path \ac{CAMEL} decoding algorithm. %

As a benchmark, we include codes designed through the \ac{QC} construction from~\cite[Sec.~IV]{SisiCamel} (C1 and C2), where we choose the number of block-rows to match the code rate of D1 and D2, respectively.  We additionally report results for the Euclidean geometry code (E5) from~\cite{SisiCamel}. As a further reference, we show the performance of the $\llbracket 800,400\rrbracket$ bicycle code (B1) decoded using the modified non-binary decoder with enhanced feedback~\cite{Babar2015,wang2012enhancedfeedback},
which has higher complexity than the \ac{CAMEL} decoding scheme. The parameters of the reference codes are summarized in Table~\ref{tab:reference_codes}.

In~\cite{SisiCamel}, an error floor attributed to the presence of numerous length-4 cycles in the Tanner graph was observed for all considered codes. In contrast, we observe from Fig.~\ref{fig:first_CAMEL} that, for code D2, plain quaternary \ac{BP} decoding exhibits no observable error floor and shows a significant waterfall behavior. In this case, \ac{CAMEL} decoding does not provide additional gains. Nevertheless, D2 achieves an error-rate performance comparable to that of the reference codes E5 and B1, despite both having lower code rates; moreover, E5 possesses a larger block length, and B1 requires higher decoding complexity.

However, for codes such as D1 and C1, where plain quaternary \ac{BP} decoding suffers from a pronounced error floor due to length-4 cycles, the \ac{CAMEL} ensemble 
decoding effectively suppresses the error floor, as shown in Fig. \ref{fig:fer_d1}. 
Consistent with the observations in~\cite{SisiCamel}, its performance closely matches that of genie-aided decoding, yielding a substantial improvement over plain quaternary \ac{BP} decoding. 
Furthermore, when using \ac{CAMEL} decoding, the 
\ac{QD} code D1 proposed in this work 
outperforms the reference code C1, despite operating at a higher rate and with a smaller block length.

Overall, the results indicate that the proposed constructions achieve high rates and competitive performance relative to state-of-the-art codes, while being paired with an effective decoding strategy capable of mitigating potential error floors.

%% file: Figures/FER_D1.tex
		\begin{tikzpicture}
		\pgfplotsset{grid style={dashed, gray}}
		\pgfplotsset{every tick label/.append style={font=\footnotesize}}
		
		\begin{axis}[%
		xshift=1.5cm,
		xmin=0.001,
		xmax=0.15,
		ymode=log,
            xmode = log,
		ymin=1e-6,
		ymax=1,
		yminorticks=true,
		axis background/.style={fill=white, mark size=1.5pt},
		xmajorgrids,
		ymajorgrids,
		width=8cm,
        height=7cm,
		xlabel={$p$},
		ylabel={LER},
		label style={font=\small},
		legend cell align={left},
		legend style={anchor = north west,  at={(0,1)}, draw=none, fill opacity=0.8, text opacity = 1,legend columns=1,font=\footnotesize, row sep = 0pt}
		]

\addplot[color=blue, line width = 0.5pt
]
table[row sep=crcr] {
0.05 1e-19\\
};

\addplot[color=KITpalegreen, line width = 0.5pt
]
table[row sep=crcr] {
0.05 1e-19\\
};

\addplot[color=black, mark=o,mark options={solid}, line width = 0.5pt
]
table[row sep=crcr] {
0.05 1e-19\\
};

\addplot[color=black, mark=triangle, dashed,mark options={solid}, line width = 0.5pt
]
table[row sep=crcr] {
0.05 1e-19\\
};

\addplot[color=black, mark=square, dotted,mark options={solid}, line width = 0.5pt
]
table[row sep=crcr] {
0.05 1e-19\\
};

\addlegendentry{New: D1 code};
\addlegendentry{C1 code};
\addlegendentry{CAMEL decoding};
\addlegendentry{Plain BP4};
\addlegendentry{Genie-aided BP4};

\addplot[
    mark=o,
color=blue,
    line width = 0.5pt
]
table[row sep=crcr] {
0.1 0.972056\\
0.09 0.940358\\
0.08 0.801047\\
0.07 0.649123\\
0.06 0.381288\\
0.05 0.174797\\
0.04 0.0504432\\
0.03 0.00594354\\
0.02 0.000305356\\
0.018 0.00015272\\
0.016 5.36207e-05\\
0.014 2.37239e-05\\
0.012 9.13681e-06\\
};

\addplot[
    mark=square,
color=blue, dotted,
    mark=square,mark options={solid},
    line width = 0.5pt
]
table[row sep=crcr] {
0.1 0.970667\\
0.0666667 0.561175\\
0.0444444 0.0921592\\
0.0296296 0.00591809\\
0.0197531 0.000251917\\
0.0131687 1.4911e-05\\
};

\addplot[
    mark=triangle,
 color=blue, dashed,
    mark options={solid},
    line width = 0.5pt
]
table[row sep=crcr] {
0.1 0.981524\\
0.0666667 0.54902\\
0.0444444 0.106555\\
0.0296296 0.0200788\\
0.0197531 0.0105184\\
0.0131687 0.00772123\\
0.00877915 0.00476146\\
0.00585277 0.00286268\\
0.00390184 0.00138653\\
0.00260123 0.00068317\\
0.00173415 0.000397046\\
0.0011561 0.000160686\\
};

\addplot[
    mark=o,
color=KITpalegreen,
    line width = 0.5pt
]
table[row sep=crcr] {
0.12 1\\
0.11 1\\
0.1 1\\
0.09 0.996875\\
0.08 0.978593\\
0.07 0.927326\\
0.06 0.825521\\
0.05 0.589118\\
0.04 0.287992\\
0.03 0.0699722\\
0.02 0.00503336\\
0.018 0.00264529\\
0.016 0.00122193\\
0.014 0.000453211\\
};

\addplot[
    mark=square,
color=KITpalegreen,
    mark=square,mark options={solid},
    line width = 0.5pt
]
table[row sep=crcr] {
0.12 1\\
0.11 1\\
0.1 1\\
0.09 0.981538\\
0.08 0.984615\\
0.07 0.935673\\
0.06 0.824289\\
0.05 0.54878\\
0.04 0.27425\\
0.03 0.0727576\\
0.02 0.00510542\\
0.018 0.00247659\\
0.016 0.00119795\\
0.014 0.000481193\\
0.012 0.000175477\\
};

\addplot[
    mark=triangle,
 color=KITpalegreen, dashed,
    mark options={solid},
    line width = 0.5pt
]
table[row sep=crcr] {
0.1 1\\
0.09 0.993769\\
0.08 0.984615\\
0.07 0.932945\\
0.06 0.822165\\
0.05 0.60728\\
0.04 0.268657\\
0.03 0.0685365\\
0.02 0.00765689\\
0.01 0.00271298\\
0.009 0.00268249\\
0.008 0.00239758\\
0.007 0.00200752\\
0.006 0.00155251\\
0.005 0.00129432\\
0.004 0.000955844\\
0.003 0.000532307\\
0.002 0.000301245\\
0.001 8.2949e-05\\
};

\end{axis}
\end{tikzpicture}

%% file: Figures/FER_codes_from_theorem2.tex
		\begin{tikzpicture}
		\pgfplotsset{grid style={dashed, gray}}
		\pgfplotsset{every tick label/.append style={font=\footnotesize}}
		
		\begin{axis}[%
		xshift=1.5cm,
		xmin=0.001,
		xmax=0.15,
		ymode=log,
            xmode = log,
		ymin=1e-6,
		ymax=1,
		yminorticks=true,
		axis background/.style={fill=white, mark size=1.5pt},
		xmajorgrids,
		ymajorgrids,
		width=8cm,
        height=7cm,
		xlabel={$p$},
		ylabel={LER},
		label style={font=\small},
		legend cell align={left},
		legend style={anchor = north west,  at={(0,1)}, draw=none, fill opacity=0.8, text opacity = 1,legend columns=1,font=\footnotesize, row sep = 0pt}
		]

\addplot[color=orange, line width = 0.5pt
]
table[row sep=crcr] {
0.05 1e-19\\
};

\addplot[color=KITorange, line width = 0.5pt
]
table[row sep=crcr] {
0.05 1e-19\\
};

\addplot[color=KITred, line width=1pt] 
table[row sep=crcr] {
0.05 1e-19\\
};

\addplot[color=black, mark=o,mark options={solid}, line width = 0.5pt
]
table[row sep=crcr] {
0.05 1e-19\\
};

\addplot[color=black, mark=triangle, dashed,mark options={solid}, line width = 0.5pt
]
table[row sep=crcr] {
0.05 1e-19\\
};

\addplot[color=black, mark=square, dotted,mark options={solid}, line width = 0.5pt
]
table[row sep=crcr] {
0.05 1e-19\\
};

\addplot[color=magenta, dashed, line width=1pt,mark=star, line width = 0.5pt
]
table[row sep=crcr] {
0.05 1e-19\\
};

\addlegendentry{New: D2 code};
\addlegendentry{C2 code};

\addlegendentry{E5}
\addlegendentry{CAMEL decoding};
\addlegendentry{Plain BP4};
\addlegendentry{Genie-aided BP4};
\addlegendentry{B1}

\addplot[
        mark=o,
    color=orange, 
    mark options={solid},
    line width = 0.5pt
]
table[row sep=crcr] {
0.1 1\\
0.09 1\\
0.08 1\\
0.07 1\\
0.06 1\\
0.05 1\\
0.04 0.895425\\
0.03 0.393878\\
0.02 0.0148191\\
0.018 0.00365212\\
0.015 0.000320347\\
0.012 9.46293e-06\\
};

\addplot[
    mark=square,
    color=orange, dotted,
    mark options={solid},
    line width = 0.5pt
]
table[row sep=crcr] {
0.15 1\\
0.135 1\\
0.12 1\\
0.105 1\\
0.09 1\\
0.075 1\\
0.06 1\\
0.045 0.983333\\
0.03 0.412961\\
0.027 0.224066\\
0.024 0.0908828\\
0.021 0.0236728\\
0.018 0.0038681\\
0.016 0.000705443\\
0.014 0.000108319\\
};

\addplot[
        mark=triangle,
    color=orange, dashed,
    mark options={solid},
    line width = 0.5pt
]
table[row sep=crcr] {
0.1 1\\
0.0666667 1\\
0.0444444 0.974118\\
0.0296296 0.354623\\
0.0197531 0.00948535\\
0.0131687 3.98889e-05\\
0.012 9.1e-06\\
0.01 3.44625e-07\\
};

\addplot[
    mark=o,
color=KITorange,
    line width = 0.5pt
]
table[row sep=crcr] {
0.1 1\\
0.09 1\\
0.08 1\\
0.07 1\\
0.06 1\\
0.05 1\\
0.04 1\\
0.03 0.995789\\
0.02 0.737977\\
0.01 0.0316307\\
0.009 0.0131923\\
0.008 0.00575085\\
0.007 0.00161755\\
0.006 0.000373464\\
0.006 0.000432269\\
};

\addplot[
    mark=square,
color=KITorange,
    mark=square,mark options={solid},
    line width = 0.5pt
]
table[row sep=crcr] {
check symplectic ok
0.1 1\\
0.09 1\\
0.08 1\\
0.07 1\\
0.06 1\\
0.05 1\\
0.04 1\\
0.03 0.997967\\
0.02 0.713821\\
0.01 0.0301439\\
0.009 0.0132138\\
0.008 0.00496552\\
0.007 0.00168969\\
0.006 0.000395624\\
};

\addplot[
    mark=triangle,
 color=KITorange, dashed,
    mark options={solid},
    line width = 0.5pt
]
table[row sep=crcr] {
check symplectic ok
0.1 1\\
0.09 1\\
0.08 1\\
0.07 1\\
0.06 1\\
0.05 1\\
0.04 1\\
0.03 0.98783\\
0.02 0.711075\\
0.01 0.0234476\\
0.009 0.0111555\\
0.008 0.0045544\\
0.007 0.00135272\\
0.006 0.000340083\\
};

\addplot [color=magenta, dashed, line width=1pt,mark=star]table[row sep=crcr]{
0.030003338898163603  0.061017519056170144\\
0.024994991652754595  0.011433569030716717\\
0.01998664440734558  0.0007844142136982535\\
0.018003338898163607  0.0002196939927297685\\
0.016  0.00005204315734195076\\
0.013956594323873122  0.000011338235012178496\\
}; 

\addplot [color=KITred, line width=1pt, mark=o]table[row sep=crcr]{
0.08 1\\
0.07 1\\
0.06 0.979592\\
0.05 0.798107\\
0.04 0.275081\\
0.03 0.0132561\\
0.03 0.0150717\\
0.028 0.00429531\\
0.026 0.00159324\\
0.024 0.000300749\\
0.02 0.000007\\
}; 
\end{axis}
\end{tikzpicture}

%% file: Sections/conclusions.tex
We presented an algebraic construction of classical and quantum \ac{QD} \ac{LDPC} codes. The resulting \ac{CSS} codes extend the family of \ac{CAMEL} codes and thereby offer increased flexibility in both block length and code rate. In terms of \ac{LER}, when stand-alone quaternary \ac{BP} decoding is limited by error-floor effects, \ac{CAMEL} ensemble decoding provides a clear performance advantage over conventional iterative decoders and achieves competitive performance to that of state-of-the-art code constructions.

%% file: main.bbl
\newpage

\begin{thebibliography}{10}
\providecommand{\url}[1]{#1}
\csname url@samestyle\endcsname
\providecommand{\newblock}{\relax}
\providecommand{\bibinfo}[2]{#2}
\providecommand{\BIBentrySTDinterwordspacing}{\spaceskip=0pt\relax}
\providecommand{\BIBentryALTinterwordstretchfactor}{4}
\providecommand{\BIBentryALTinterwordspacing}{\spaceskip=\fontdimen2\font plus
\BIBentryALTinterwordstretchfactor\fontdimen3\font minus \fontdimen4\font\relax}
\providecommand{\BIBforeignlanguage}[2]{{%
\expandafter\ifx\csname l@#1\endcsname\relax
\typeout{** WARNING: IEEEtran.bst: No hyphenation pattern has been}%
\typeout{** loaded for the language `#1'. Using the pattern for}%
\typeout{** the default language instead.}%
\else
\language=\csname l@#1\endcsname
\fi
#2}}
\providecommand{\BIBdecl}{\relax}
\BIBdecl

\bibitem{Shor1995}
P.~W. Shor, ``Scheme for reducing decoherence in quantum computer memory,'' \emph{Physical Review A}, vol.~52, no.~4, pp. R2493--R2496, 1995.

\bibitem{Steane1996}
A.~M. Steane, ``Error correcting codes in quantum theory,'' \emph{Physical Review Letters}, vol.~77, no.~5, pp. 793--797, 1996.

\bibitem{Calderbank1996}
A.~R. Calderbank and P.~W. Shor, ``Good quantum error-correcting codes exist,'' \emph{Physical Review A}, vol.~54, no.~2, pp. 1098--1105, 1996.

\bibitem{Gottesman1997}
\BIBentryALTinterwordspacing
D.~Gottesman, ``Stabilizer codes and quantum error correction,'' Ph.D. dissertation, California Institute of Technology, 1997. [Online]. Available: \url{https://arxiv.org/abs/quant-ph/9705052}
\BIBentrySTDinterwordspacing

\bibitem{Nielsen2010}
M.~A. Nielsen and I.~L. Chuang, \emph{Quantum Computation and Quantum Information}.\hskip 1em plus 0.5em minus 0.4em\relax Cambridge, UK: 10th Anniversary Edition, Cambridge University Press, 2010.

\bibitem{calderbank1998quantum}
A.~R. Calderbank, E.~M. Rains, P.~W. Shor, and N.~J.~A. Sloane, ``Quantum error correction via codes over {GF}(4),'' \emph{IEEE Transactions on Information Theory}, vol.~44, no.~4, 1998.

\bibitem{Gallager}
R.~G. Gallager, ``Low-density parity-check codes,'' \emph{IRE Trans. Inf. Theory}, vol. IT-8, no.~1, pp. 21--28, 1962.

\bibitem{MacKay1999}
D.~J.~C. MacKay, ``Good error-correcting codes based on very sparse matrices,'' \emph{IEEE Trans. Inf. Theory}, vol.~45, pp. 399--432, 1999.

\bibitem{Poulin2008}
D.~Poulin and Y.~Chung, ``On the iterative decoding of sparse quantum codes,'' \emph{Quantum Information and Computation}, vol.~8, no.~10, pp. 987--1000, 2008.

\bibitem{Babar2015}
Z.~Babar, P.~Botsinis, D.~Alanis, S.~X. Ng, and L.~Hanzo, ``{Fifteen years of quantum LDPC coding and improved decoding strategies},'' \emph{IEEE Access}, vol.~3, pp. 2492--2519, 2015.

\bibitem{Hagiwara2007}
M.~Hagiwara and H.~Imai, ``{Quantum quasi-cyclic LDPC codes},'' in \emph{Proc. IEEE International Symposium on Information Theory (ISIT)}, 2007, pp. 806--810.

\bibitem{Tang2024}
\BIBentryALTinterwordspacing
C.~Tang, C.~Bai, and Y.~Feng, ``New quantum {LDPC} codes based on projective geometry,'' \emph{Scientific Reports}, vol.~14, no.~1, p. 17014, 2024. [Online]. Available: \url{https://doi.org/10.1038/s41598-024-67786-0}
\BIBentrySTDinterwordspacing

\bibitem{Kasai2025}
K.~Kasai, ``Quantum error correction with girth-16 non-binary {LDPC} codes via affine permutation construction,'' in \emph{Proc. International Symposium on Topics in Coding (ISTC)}, 2025, pp. 1--5.

\bibitem{SisiCamel}
S.~Miao, J.~Mandelbaum, H.~Jäkel, and L.~Schmalen, ``A joint code and belief propagation decoder design for quantum {LDPC} codes,'' in \emph{Proc. IEEE International Symposium on Information Theory (ISIT)}, 2024, pp. 2263--2268.

\bibitem{baldelli_ISTC_2025}
A.~Baldelli, M.~Battaglioni, and P.~Santini, ``Quantum {CSS} {LDPC} codes with quasi-dyadic structure,'' in \emph{Proc. International Symposium on Topics in Coding (ISTC)}, 2025, pp. 1--5.

\bibitem{Martinez2022}
M.~Martinez and C.~A. Kelley, ``Minimum distance and other properties of quasi-dyadic parity check codes,'' in \emph{Proc. IEEE International Symposium on Information Theory (ISIT)}, 2022, pp. 2118--2123.

\bibitem{dyadics}
M.~Martinez, T.~Pllaha, and C.~A. Kelley, ``Codes based on dyadic matrices and their generalizations,'' \emph{Advances in Mathematics of Communications}, vol.~19, no.~5, pp. 1277--1300, 2025.

\bibitem{santini2022reproducible}
P.~Santini, E.~Persichetti, and M.~Baldi, ``Reproducible families of codes and cryptographic applications,'' \emph{Journal of Mathematical Cryptology}, vol.~16, no.~1, pp. 20--48, 2022.

\bibitem{Tanner1981}
M.~R. Tanner, ``A recursive approach to low complexity codes,'' \emph{IEEE Transactions on Information Theory}, vol.~27, no.~5, pp. 533--547, Sep. 1981.

\bibitem{pacenti_margulis}
\BIBentryALTinterwordspacing
M.~Pacenti, D.~Chytas, and B.~Vasic, ``Construction and decoding of quantum {M}argulis codes,'' 2025. [Online]. Available: \url{https://arxiv.org/abs/2503.03936}
\BIBentrySTDinterwordspacing

\bibitem{wang2012enhancedfeedback}
Y.-J. Wang, B.~C. Sanders, B.-M. Bai, and X.-M. Wang, ``Enhanced feedback iterative decoding of sparse quantum codes,'' \emph{IEEE Transactions on Information Theory}, vol.~58, no.~2, 2012.

\end{thebibliography}
